\documentclass[aps]{revtex4-1}
\usepackage{amsmath}
\usepackage{latexsym}
\usepackage{float} 
\usepackage{amssymb} 
\usepackage{graphicx} 
\usepackage{epsfig}

\begin{document}

\title{Sterically stabilized lock and key colloids: A self-consistent field 
theory study}
\author{S. A. Egorov\footnote{Corresponding author e-mail: 
sae6z@virginia.edu}}
\affiliation{Department of Chemistry, University of 
Virginia, McCormick Road, Charlottesville, Virginia 22904, USA}
\begin{abstract}
\begin{center}
Abstract
\end{center}
A self-consistent field theory study of lock and key type
interactions between sterically stabilized colloids in polymer
solution is performed. 
 Both the key particle and the lock cavity are assumed to have
 cylindrical shape, and their surfaces are uniformly grafted with
 polymer chains. The lock-key potential of mean force is computed for 
 various model parameters, such as length of free and grafted chains, lock
 and key size matching, free chain volume fraction, grafting density, 
and various enthalpic interactions present in the
 system. The lock-key interaction is found to be highly tunable, which
 is important in the rapidly developing field of particle self-assembly.  
\end{abstract}
\maketitle
\newcommand{\nc}{\newcommand}
\nc{\cntr}[1]{\begin{center}{\bf #1}\end{center}}
\nc{\ul}{\underline} 
\nc{\fn}{\footnote}
\nc{\pref}{\protect\ref}
\setlength{\parindent}{.25in}
\nc{\stab}[1]{\begin{tabular}[c]{c}#1\end{tabular}}
\nc{\doeps}[3]{\stab{\setlength{\epsfxsize}{#1}\setlength{\epsfysize}{#2}
\epsffile{#3}}}
\nc{\sq}{^{2}}
\nc{\eps}{\epsilon}
\nc{\sig}{\sigma}
\nc{\lmb}{\lambda}
\nc{\lbr}{\left[}
\nc{\rbr}{\right]}
\nc{\lpar}{\left(} 
\nc{\rpar}{\right)}
\nc{\ra}{\rightarrow}
\nc{\Ra}{\Rightarrow}
\nc{\lra}{\longrightarrow}
\nc{\LRa}{\Longrightarrow}
\nc{\la}{\leftarrow}
\nc{\La}{\Leftarrow}
\nc{\lla}{\longleftarrow}
\nc{\LLa}{\Longleftarrow}
\nc{\EE}[1]{\times 10^{#1}}
\nc{\simleq}{\stackrel{\displaystyle <}{\sim}}
\newcommand{\be}{\begin{equation}}
\newcommand{\ee}{\end{equation}}
\newcommand{\bea}{\begin{eqnarray}}
\newcommand{\eea}{\end{eqnarray}}
\newcommand{\R}{\vec{R}}
\newcommand{\bc}{\begin{center}}
\newcommand{\ec}{\end{center}}
%\clearpage

\section{Introduction}
\label{sc1}

Self-assembly of colloidal particles into desired structures, which is driven 
by selective and directional interparticle interactions, can provide a 
promising route for fabrication of novel 
materials.\cite{kumar10}
One particularly promising approach in this research area is based on 
utilizing  ``lock and key'' colloidal systems.\cite{sacanna10,solomon10} 
As such, these model systems have recently received substantial
attention both experimentally\cite{sacanna10,solomon10}  and 
theoretically.\cite{kinoshita02,silvestre04,konig08,odriozola08,jin11}

In a recent experimental study,\cite{sacanna10} an efficient method
has been developed to produce colloidal lock particles containing a
spherically shaped cavity. The interaction between these lock
particles and complementary spherical key particles is comprised
of two major contributions: Coulomb repulsion arising due to charge
stabilization, and depletion attraction\cite{asakura54} due to the
presence of the polymeric depletants in solution. The depletion attraction
is the strongest when the radius of the key particle exactly matches
that of the cavity due to the fact that the overlap of excluded
volumes associated with lock and key particles is maximized in this
case. As a result, this key-lock binding is highly specific, which is
one of the crucial requirements for developing a successful
self-assembly process. 
In addition to geometric considerations, 
the strength of depletion attraction can be controlled either by
varying the depletant concentration or by adjusting temperature, 
thereby changing the polymeric depletant size.\cite{sacanna10} 
Hence, the key-lock
interaction is not only selective, but also reversible and tunable,
which is equally important for controlling self-assembly. In this
regard, it was found that the binding-unbinding transition in the
key-lock system is significantly sharper 
when steric stabilization (arising due to the presence of grafted
polymer layers on both key and lock surfaces) is used instead of
charge stabilization.\cite{sacanna10} The sharpness of 
the binding-unbinding transition is determined by considering 
the fraction of occupied lock cavities as a function of depletant
concentration. Experimental data show that for charge stabilized
key-lock systems this function grows gradually, while for sterically
stabilized systems it approaches a step function, i.e. the
binding-unbinding transition becomes much sharper.\cite{sacanna10} 

On the theoretical side, key-lock interactions in colloidal systems have
been studied using integral equation theory,\cite{kinoshita02}, 
density functional theory,\cite{konig08} and molecular simulation
techniques.\cite{odriozola08} Very recently, a highly efficient
hybrid approach to study key-lock  model systems 
(among other multidimensional problems) 
has been developed, which combines Monte Carlo
simulation for obtaining microscopic configurations of the depletant
particles and density functional theory for calculating the free
energy.\cite{jin11} All these previous studies employed microscopic models
based exclusively on hard-sphere excluded volume interactions, whereby
the depletion attraction between lock and key particles is driven by
the entropic solvent contribution. In agreement with experimental
observations and geometric arguments based on excluded volume overlaps,
it has been found that the strongest binding interaction occurs for
the systems where the size of the key exactly matches the size of the
lock cavity. However, it was also found that a simple Asakura-Oosawa
type treatment\cite{asakura54} based exclusively on considering excluded volume
overlaps is not always sufficient.
In particular, these studies have highlighted the
importance of employing a realistic microscopic model for the solvent
(depletant) by showing that the Asakura-Oosawa
approach, which assumes ideality of the solvent,
can significantly underestimate the selectivity of key-lock 
interactions.\cite{kinoshita02}  

While earlier theoretical studies of lock and key model systems have
considered several different geometric shapes of key particles and lock
cavities, the depletant has always been treated as a simple hard
sphere solvent, thereby limiting the key-lock interaction to purely
entropic depletion attraction. At the same time, in the experimental
work the strength of key-lock binding has been tuned by changing the
effective size of polymeric depletant via temperature changes, which
indicates the potential importance of enthalpic effects in these
systems.\cite{sacanna10} Furthermore, the fact  that the sharpness of
binding-unbinding transition can be increased in the presence of
steric stabilization,\cite{sacanna10}
points to the necessity of developing a microscopic model of lock and
key systems that would explicitly include grafted chains  
on the surfaces of lock and key particles. 
The central goal of the present work is to develop such a model. 

In our earlier work, we have employed mean-field type theoretical
methods (density functional theory and self-consistent field (SCF)
theory\cite{fleer93}) 
to study interactions between sterically stabilized colloidal
particles in polymer melts and solutions.\cite{egorov07,egorov10b} 
By comparing theoretical
results with computer simulations, it was shown that theory yields
accurate density profiles of both free and
grafted chains,\cite{egorov10b} 
as well as potentials of mean force (PMF) between sterically
 stabilized colloids.\cite{egorov07} 
In the present work, we employ mean-field
techniques to study interactions between sterically stabilized lock
and key particles in the presence of polymeric depletants. 
We compute the corresponding PMFs and study their dependence on the 
geometric matching between the sizes of key particle and
lock cavity, which is the crucial parameter governing the strength of
lock-key interactions.\cite{sacanna10} 
In addition, we study the dependence of PMF on
several other model parameters, which are known to play an important
role in governing the microstructure of sterically stabilized 
colloids in polymer melts and solutions, 
such as the ratio of free and grafted chain 
lengths,\cite{kumar10,dutta08,green06,bansal05,bansal06}
grafting density,\cite{kumar10,dutta08} 
and enthalpic interactions of free chains with
grafted chains\cite{egorov10b} and with the surface of lock and key
particles.       

The remainder of the paper is organized as follows. In
Section~\ref{sc2} we introduce our microscopic model and describe
SCF-based approach for calculating PMF between sterically stabilized
lock and key particles. The results of our calculations of PMFs for a
wide range of model parameters are presented in Section~\ref{sc3}.
Section~\ref{sc4} concludes the paper.

\section{Microscopic Model and Theory}
\label{sc2}

\begin{figure}
\includegraphics[width=10cm,angle=0]{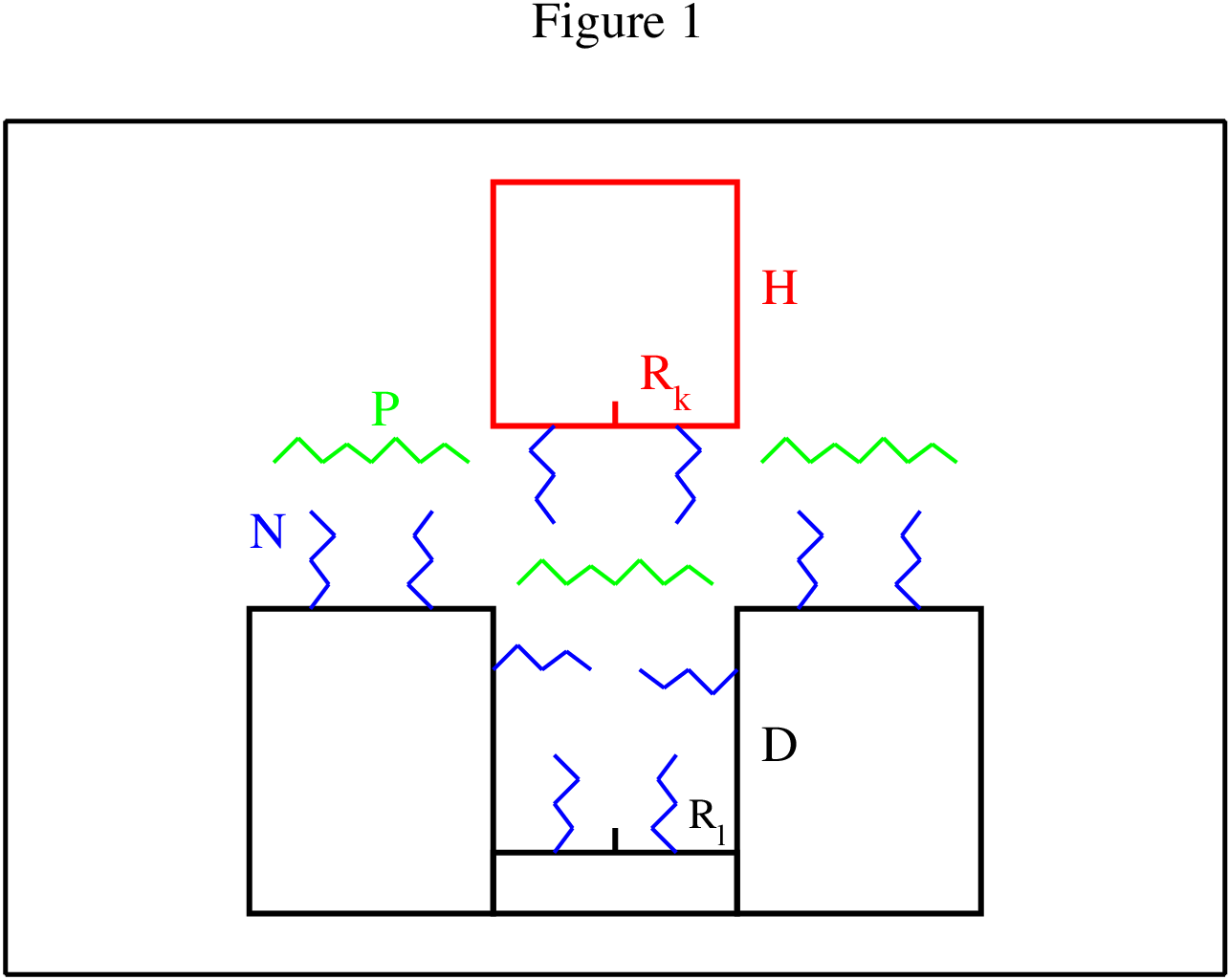}
%\hspace{0.50cm}
%\includegraphics[width=8.5cm,angle=0]{EPS/snap2_c.eps}
%\vspace{0.8cm} 
\caption{A schematic side-view of the microscopic model. The key
  particle (shown in red) is a cylinder of height $H$ and radius
  $R_k$; the lock cavity (shown in black) is a cylinder of height $D$
  and radius $R_l$. Both particles are uniformly grafted with polymer
  chains of length $N$ (shown in blue) at grafting density $\sigma_g$ 
  and immersed in a solution of
  polymer chains of length $P$ (shown in green) with bulk volume
  fraction $\phi_b$.}
\label{fig1} 
\end{figure}

Previous studies of lock-key interactions have considered various
geometries for the particles, including 
spherical and ellipsoidal
shapes.\cite{kinoshita02,silvestre04,konig08,odriozola08,jin11} In the
present work, we assume cylindrical shape both for the key particle
and for the lock cavity. In particular, the key particle is assumed
to be a cylinder of radius $R_k$ and height $H$, while the lock cavity
is characterized by radius $R_l$ and depth $D$. The conditions
$R_k=R_l$ and $H=D$ correspond to the perfect size matching between
lock and key. The particles are immersed in a solution of polymer
chains of length $P$ with bulk volume fraction $\phi_b$. In addition,
the surfaces of the particles are uniformly grafted with chains of
length $N$ at grafting density $\sigma_g$. In what follows, we
consider only coaxial arrangements of the cylinders describing key
particle and lock cavity, both of which are aligned along the $z$ axis. 
As a result, the lock-key PMF is a function of a single coordinate
$z$. A schematic side-view of our microscopic model is presented in
Fig.~\ref{fig1} 
 
We use SCF theory to obtain the PMF between
sterically stabilized lock and key particles in a polymer solution.
The central quantity in the SCF approach is mean-field free energy, which is 
expressed as a functional of the volume fraction profiles and SCF potentials 
for all components in the system. Minimization of this free energy under the 
incompressibility constraint and appropriate boundary conditions yields 
equilibrium density distributions of various components. For most problems of 
interest, SCF equations need to be solved iteratively and numerically, which 
necessarily involves space discretization, i.e. use of a lattice. 
Here we employ the method of Scheutjens and 
Fleer,\cite{fleer93} which uses the segment diameter $\sigma$ 
as the size of the cell; throughout this work all the distances are reported 
in units of the cell size $\sigma$.

For the present problem, the SCF equations are set up on a cylindrical
lattice and solved using two-gradient approach, where the 
density profiles vary both radially and
laterally.\cite{huinink97,steels00,steels00b} 
The lattice layers along the axis of the cylinder are numbered
according to $z=1,2,\cdots,N_z$, while circular arrangements of the
lattice sites within each $z$-layer are numbered as
$r=1,2,\cdots,N_r$.  Due to the azimuthal symmetry of the problem,
there are $L(r)=\pi[r^2-(r-1)^2]$ indistinguishable sites at each
coordinate $(z,r)$. In order to implement SCF formalism, one needs to
define {\em a priori} step probabilities, which are determined by the
fractions of sites in neighboring layers adjacent to a given site on a
lattice. In the $r$-direction, these probabilities follow from
geometric considerations:\cite{huinink97}
\bea
\Lambda(r|r-1)&=&\frac{S(r-1)}{3L(r)},\\ \nonumber
\Lambda(r|r+1)&=&\frac{S(r)}{3L(r)},\\ \nonumber
\Lambda(r|r)&=& 1-\Lambda(r|r-1)-\Lambda(r|r+1),
\label{lambdar}
\eea
where $S(r)=2\pi r$ is the surface area between the two cylinders of
unit height and radii $r$ and $r+1$. The transition probabilities
along the $z$-direction are given by:\cite{huinink97} 
$\Lambda(z,z^{\prime})=1/3$, where $z^{\prime}=z-1,z,z+1$. Overall, for each
lattice site $(z,r)$, there are nine transition probabilities given
by:
\be
\lambda(z,r|z^{\prime},r^{\prime})=\Lambda(z,z^{\prime})\Lambda(r,r^{\prime})
=\frac{1}{3}\Lambda(r,r^{\prime}),
\label{lambdazr}
\ee
where $(z^{\prime},r^{\prime})=(z+\alpha,r+\beta)$, with $\alpha,\beta=-1,0,1$. 
The transition probabilities thus defined are properly normalized, 
with their sum equal to unity.

In our model, both lock and key particle are assumed to be uniformly
composed of a fixed number of segments of type $p$ that fill the corresponding
geometrical forms discussed above. Accordingly, the particles are
impenetrable to any other segment types present in the system, i.e. 
the volume fraction $\phi_p(z,r)$=1 
for all the lattice sites located inside the particles
and $\phi_p(z,r)$=0 otherwise. In addition to the particles, there are
three other segment types in our model, those of free chains
($f$), grafted chains ($g$), and solvent ($w$). The latter fill all
the remaining vacancies on the lattice, thereby ensuring the
incompressibility condition. 

The SCF equation for the volume fraction 
of grafted chains is written in terms of
propagators as follows:\cite{fleer93,huinink97,steels00,steels00b} 
\be
\phi_g(z,r)=C_{g}\sum_{s=1}^{N}\frac{G_g(z,r,s|1)G_g(z,r,s|N)}{G_g(z,r)},
\label{phig}
\ee
where the sum runs over all the segments of the grafted chain, the
normalization constant $C_g$ is determined by the grafting density 
$\sigma_g$, and $G_g(z,r)$ is given in terms of the 
potential  $u_g(z,r)$:
\be
G_g(z,r)=\exp(-\beta u_g(z,r)),
\label{ugzr}
\ee
where $\beta=1/k_BT$. 

The potential $u_g(z,r)$ is comprised of two terms:
\be
u_g(z,r)=u^{hc}(z,r)+u_{g}^{int}(z,r),
\label{ugsum}
\ee
where the hard-core potential $u^{hc}(z,r)$ is independent of the
segment type and serves as a Lagrange multiplier enforcing the
incompressibility condition, while the interaction potential is given
by:
\be
\beta u_{g}^{int}(z,r)=\frac{1}{2}\sum_{i\neq g}\chi_{ig}<\phi_{i}(z,r)>,
\label{ugint}
\ee
where the sum runs over all the segment types other than $g$, 
$\chi_ {ig}$ are the corresponding Flory-Huggins interaction parameters, 
and the step-weighted volume fractions are defined as follows:
\be
<\phi_{i}(z,r)>=\sum_{z^{\prime}}\sum_{r{^\prime}}\lambda(z,r|z^{\prime},r^{\prime})
\phi_{i}(z^{\prime},r^{\prime}).
\label{phiav}
\ee
As our boundary condition for the grafted chains, 
we pin their end-segments to the surfaces of lock and key particles as
depicted in Fig.~\ref{fig1}.\cite{fleer93,steels00,steels00b}  

The SCF equation for the volume fraction 
of free chains is similar to Eq.~(\ref{phig}), except that the
normalization constant is obtained from the value of the bulk volume
fraction of free chains $\phi_b$ (instead of the grafting density),
and no pinning of the end-segments is performed.  
Finally, the equation for the solvent is straightforward:
\be
\phi_{w}(z,r)=\exp(-\beta u_{w}(z,r)).
\label{phiw}
\ee
By imposing the incompressibility 
constraint, we solve SCF equations simultaneously and iteratively to
obtain the equilibrium 
volume fractions of all the components present in the system. 

From the resulting volume fractions, one can compute the Helmholtz
free energy of the system as follows:\cite{fleer93,steels00,steels00b}  
\be
\beta A = \sum_{i=1}^{m}\sum_{z=1}^{N_{z}}\sum_{r=1}^{N_{r}}
L(r)\phi_{i}(z,r)\left\{\frac{\ln \phi_{b}^{i}}{N_{i}}+
\ln G_{i}(z,r)+\sum_{j>i}^{m}\chi_{ij}<\phi_{j}(z,r)>\right\},
\label{helm}
\ee
where the summation index $i$ runs over all the segment types present
in the system; $N_{i}=$ 1, $N$, and $P$ for $i=w,g,$ and $f$,
respectively; $\phi_{b}^{f}$ is equal to the bulk volume fraction
$\phi_b$ for the free chains, while for the grafted chains
\be
\phi_{b}^{g}=\frac{nN}{\sum_{z=1}^{N_{z}}\sum_{r=1}^{N_{r}}L(r)G_{g}(z,r,N|1)},
\label{phibg}
\ee
where $n$ is the number of grafted molecules in the system determined
by the grafting density.
We note that the total Helmholtz energy can be split into
the contribution due to the grafted chains, $\beta A_{g}$ (given by the
term $i=g$ in the sum over $i$ in Eq.~(\ref{helm})), and the
contribution due to the mobile species, i.e. free chains and the
solvent,  $\beta A_{wf}$ (given by the
terms $i=w,f$ in the sum over $i$ in
Eq.~(\ref{helm})).\cite{vanlent90} 

The dimensionless lock-key PMF as a function of lock-key separation
$z$ is given by the difference of the Helmholtz free energy at the
particle separation $z$ and at infinite separation. 
\be
\beta W(z)=\beta A(z)-\beta A(\infty).
\label{pmf}
\ee
By splitting the free energy into contributions from grafted and
mobile species, the PMF can be decomposed into the term due to grafted
chains and the term due to mobile species:\cite{dutta08,vanlent90} 
\be
\beta W(z)=\beta W_{g}(z)+\beta W_{wf}(z).
\label{pmfsplit}
\ee

Note that here we focus exclusively on the polymer- and
solvent-mediated part of the PMF, i.e. we do not consider the bare
interaction between lock and key particles, the latter term can always
be added separately to the PMF.

\section{Results}
\label{sc3}

As discussed in the Introduction, prior experimental studies of lock
and key colloids have shown that the sharpness of
binding-unbinding transition can be increased by grafting polymeric 
chains on the surfaces of lock and key particles.\cite{sacanna10}
Furthermore, recent experimental results in the field of polymer
nanocomposites indicate that the interaction between sterically
stabilized colloids in polymer melts and solutions can be tuned from
repulsive to attractive by varying the ratio of free and grafted chain 
lengths.\cite{kumar10,dutta08,green06,bansal05,bansal06} 
Hence, we begin by studying the effect of the ratio $P/N$ on the PMF
between sterically stabilized lock and key particles.

\begin{figure}
\includegraphics[width=10cm,angle=0]{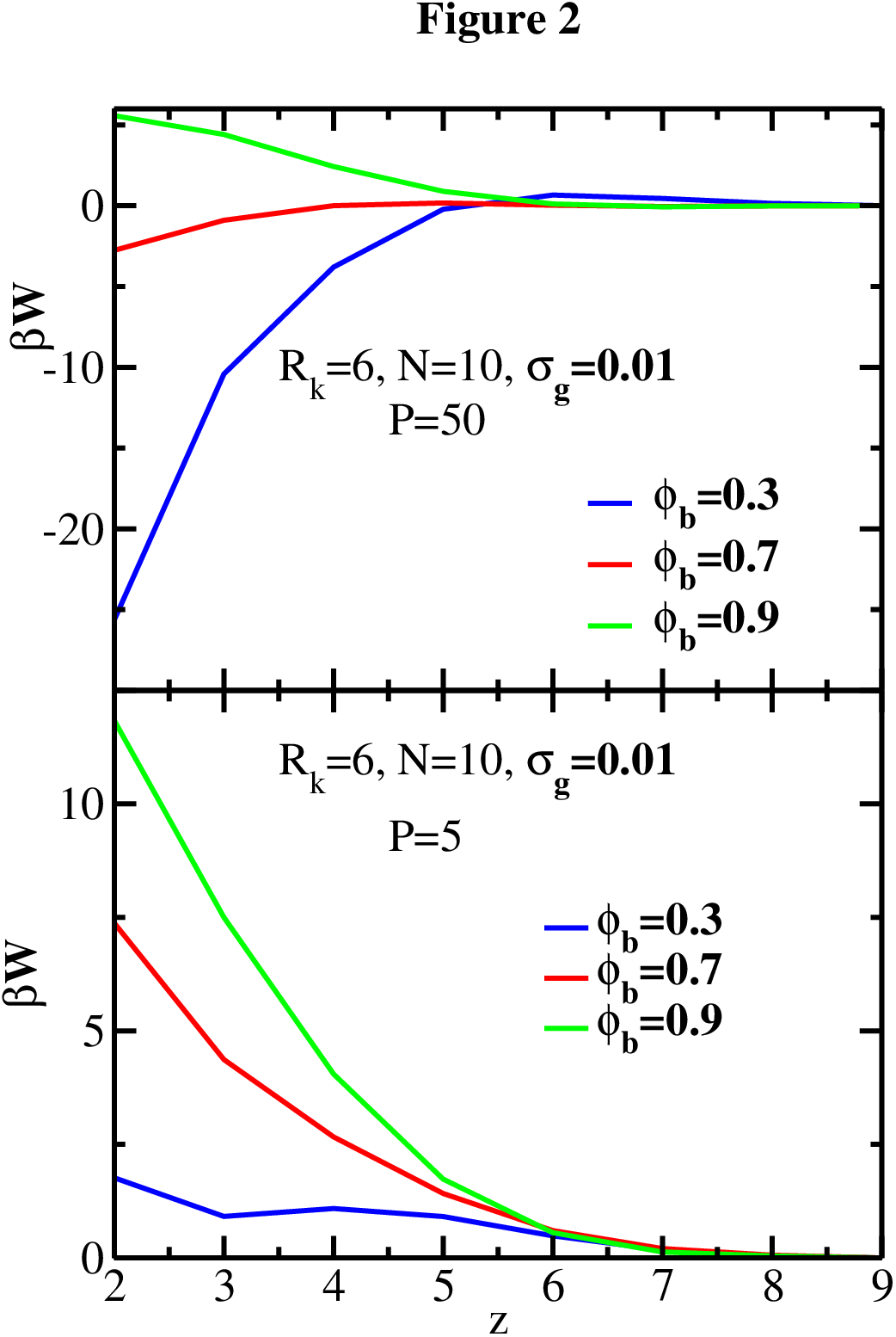}
%\hspace{0.50cm}
%\includegraphics[width=8.5cm,angle=0]{EPS/snap2_c.eps}
%\vspace{0.8cm} 
\caption{The lock-key PMF between a cylindrical key of radius $R_k=6$
  and a cylindrical cavity of radius $R_l=6$ at three different values
of the free polymer bulk volume fraction. The length of the grafted
chains is $N=10$ and their grafting density is $\sigma_g=0.01$. Upper
panel: the length of free chains is $P=50$, lower panel: $P=5$.}
\label{fig2} 
\end{figure}

The corresponding results are presented in Fig.~\ref{fig2}, where we plot
dimensionless PMF as a function of the lock-key separation for several
values of the free polymer volume fraction. In our lattice model,  
the surface of the bottom of the lock cavity corresponds to the surface layer 
$z=1$, and the lock-key separation is measured from the leading edge of the 
key particle, i.e. $z=2$ on the graph corresponds to lock and key being in 
contact. 
Throughout this study, we fix 
the height of the key and the depth of the lock at $H=D=4$, 
the radius of the lock cavity at $R_l$=6, and 
the length of the grafted chains at $N=10$.
In the upper panel, we set  
$P/N=5$, and in the lower panel, $P/N=0.5$. In both cases, the
grafting density is fixed at $\sigma_g=0.01$ and the key particle radius is
taken to be $R_k=6$, i.e. the lock and key sizes match perfectly. 
One sees that in the case when the length of the free chains is
smaller than that of the grafted chains, the polymer-induced component 
is repulsive for all values of the free polymer volume fraction
considered, which is similar to the behavior of sterically stabilized
colloidal particles.\cite{kumar10,dutta08,green06,bansal05,bansal06} 
By decomposing the total PMF into contributions due to mobile species
and grafted polymers (not shown), one finds that the
repulsion is primarily due to  the overlap of the grafted chains,
which outweighs the attractive term coming from the osmotic pressure
of the mobile species in solution (again in analogy to
 sterically stabilized colloids\cite{dutta08}). 

\begin{figure}
\includegraphics[width=10cm,angle=0]{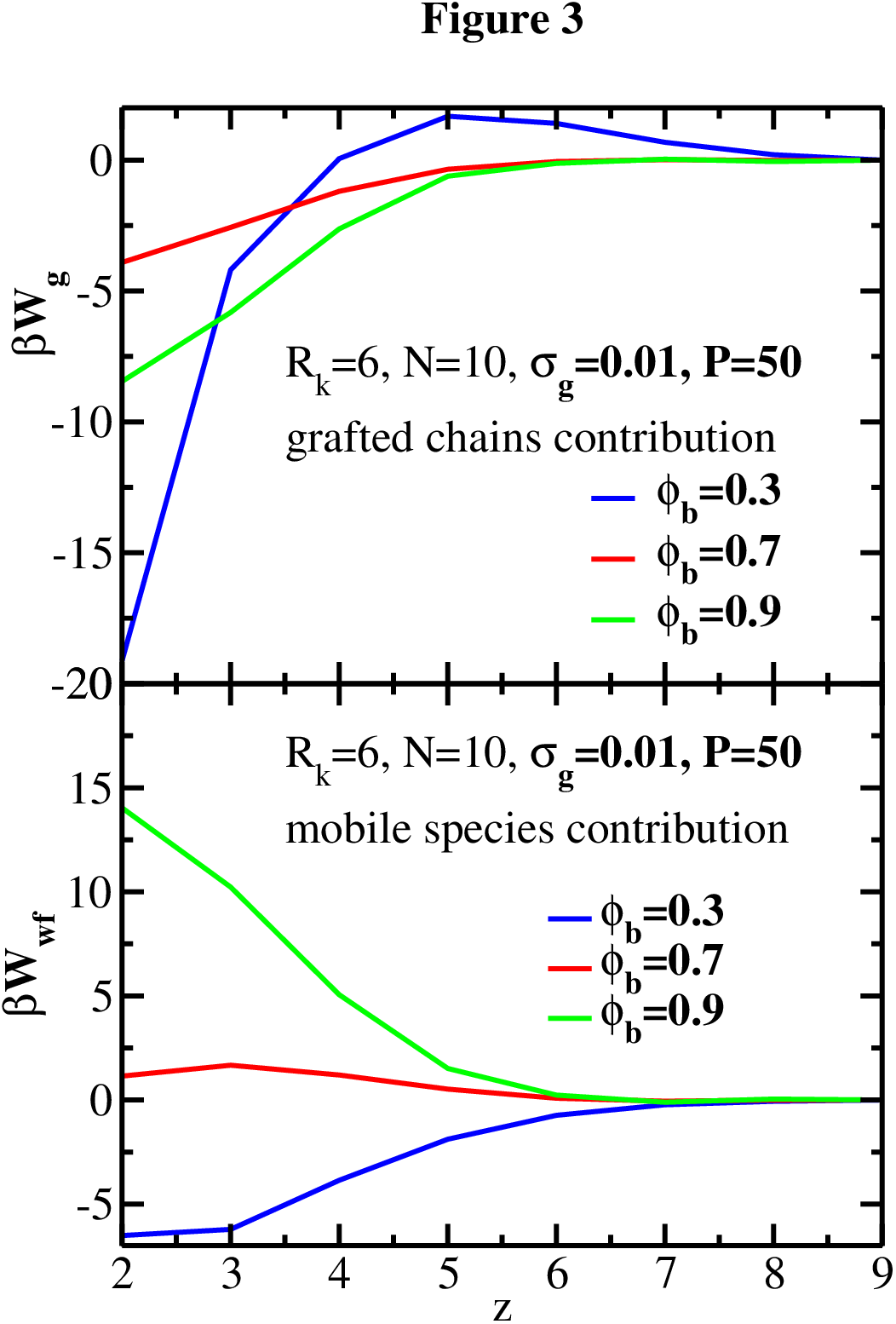}
\caption{The decomposition of lock-key PMF shown in the upper panel of
Fig.~2 into contributions due to grafted chains (upper panel) and
mobile species (lower panel).}
\label{fig3} 
\end{figure}

On the other hand, when free chains are longer than the grafted ones, 
the polymer mediated PMF is strongly attractive at low polymer volume
fraction ($\phi_b$=0.3), moderately attractive at intermediate volume
fraction ($\phi_b$=0.7), and repulsive at high volume fraction
($\phi_b$=0.9). We note here that in the case of flat polymer brushes, 
the phenomenon of autophobic dewetting would lead to an attractive PMF in the 
regime $P/N>1$ for all values of free polymer volume
fractions.\cite{kumar10,dutta08,green06,bansal05,bansal06}  
However, for highly curved brushes (grafted on spherical or
cylindrical surfaces), melt-brush interpenetration and wetting is
strongly enhanced. Indeed, recent SCF calculations have shown that
wetting-dewetting transition is shifted to much larger values of
$P/N$ ratio for highly curved sterically stabilized
nanoparticles.\cite{trombly10} 

This change in the nature of the lock-key PMF
from attractive to repulsive with increasing volume fraction (when
$P>N$) provides one possible way for tuning interactions in
sterically stabilized lock-key colloidal systems. 
Hence, it would be of interest to decompose the total
polymer-induced PMF according to Eq.~(\ref{pmfsplit}) 
into two separate contributions arising due to
mobile species and grafted chains. This decomposition is shown in
Fig.~\ref{fig3}, where  the contribution due to grafted chains is
given in the upper panel, and the contribution due to mobile species in
the lower panel. Compared to the case when $P<N$, the situation is
reversed in that the term due to grafted chains is predominantly
attractive, except for a weak repulsive barrier at intermediate
lock-key separations at the lowest volume fraction considered. The
term coming from osmotic pressure due to mobile species is attractive at
the low volume fraction, weakly repulsive at the intermediate
$\phi_b$, and strongly repulsive at the highest value of $\phi_b$
studied. Once again, this behavior is qualitatively 
similar to sterically stabilized
colloids, where it was rationalized by arguing that mixing of grafted
chains from two particles is more favorable than mixing of short
grafted chains and long free chains.\cite{dutta08}

\begin{figure}
\includegraphics[width=10cm,angle=0]{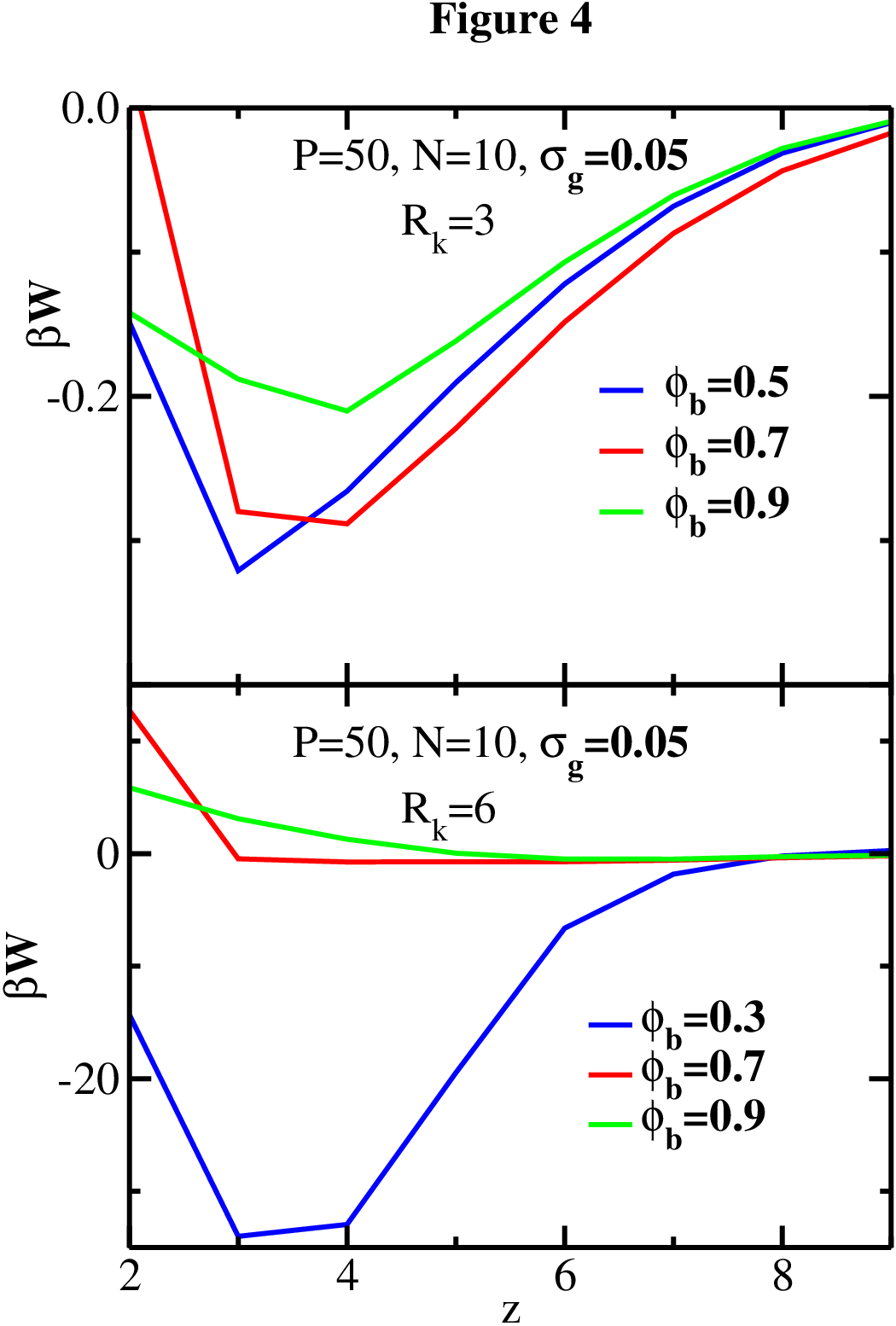}
\caption{The lock-key PMF for a cylindrical cavity of radius $R_l$=6
  and for two values of the key particle radius: $R_k$=3 (upper panel)
  and $R_k$=6 (lower panel).}
\label{fig4} 
\end{figure}
 
As a next step, we study the effect of the size matching between the
lock and key particles on the PMF. To this end, we set the grafting
density at $\sigma_g=0.05$, the length of the free chains at $P=50$,
and compute the lock-key PMF for two sizes of the key particle: 
$R_k=6$ (perfect size matching), and $R_k=3$ (key is half the size of
the lock cavity). Our results are shown in the
upper panel of Fig.~\ref{fig4} for the smaller key, and in the lower
panel for the larger key.  First of all, one immediately notices 
that the lock-key
interaction is substantially weaker in the absence of perfect
geometric matching, in agreement with earlier theoretical and
simulation studies.\cite{kinoshita02,odriozola08,jin11}
In addition, while the lock-key PMF for the
smaller key is weakly attractive at all
values of free polymer volume fraction, for the larger 
key it changes from strongly attractive at low volume
fraction to repulsive at higher values of $\phi_b$, similarly to what
was seen earlier in the upper panel of Fig.~\ref{fig2}. 
By comparing the lower panel of Fig.~\ref{fig4} with the 
upper panel of Fig.~\ref{fig2}, one can see the effect of the grafting
density on the lock-key PMF (all other model parameters being equal in
these two panels). The most pronounced difference is the abrupt change
in the slope of the PMF at short lock-key separations in the case of
higher grafting density, which leads to a well-defined minimum in the
PMF at the low free polymer volume fraction.

\begin{figure}
\includegraphics[width=10cm,angle=0]{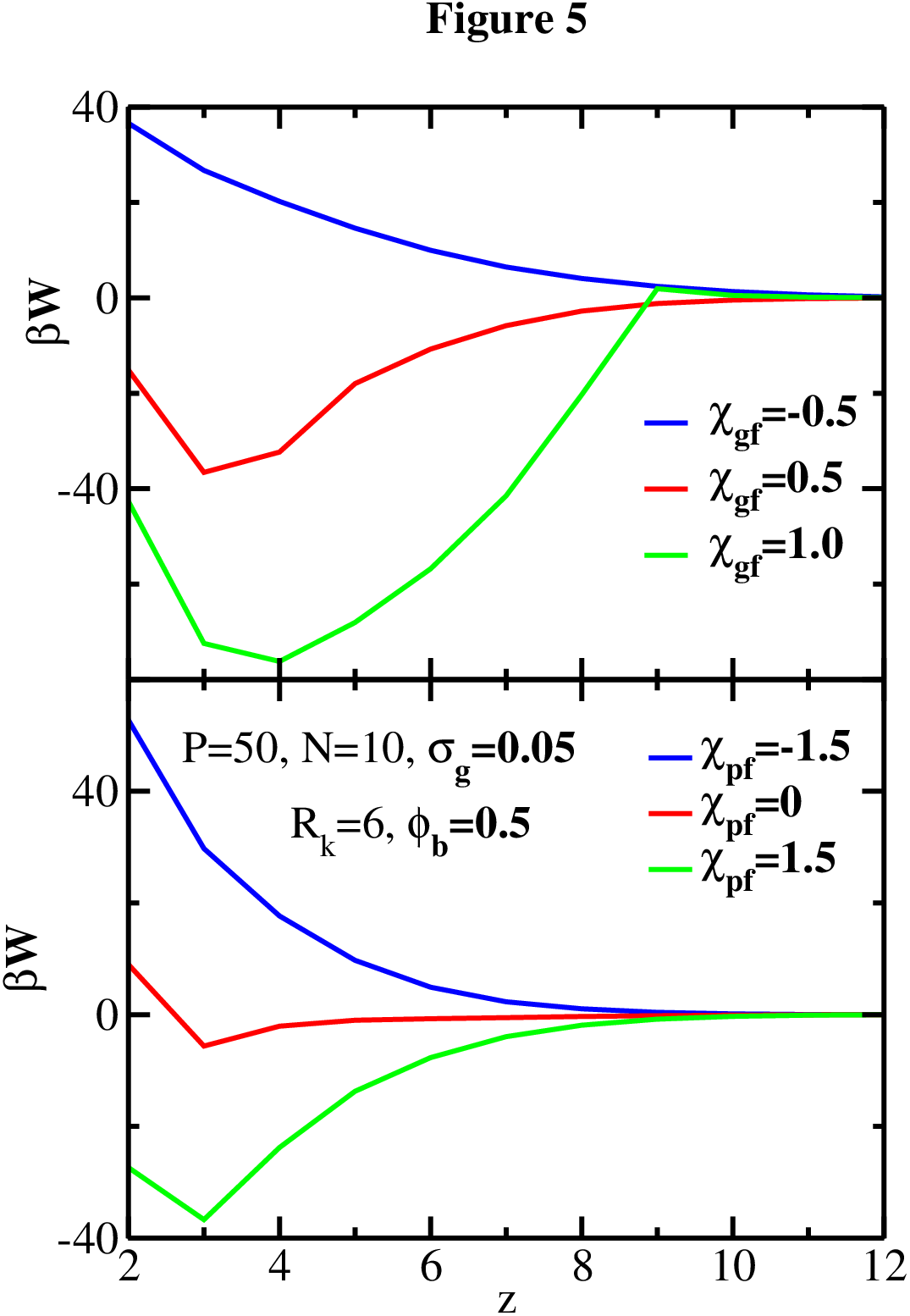}
\caption{The lock-key PMF between a cylindrical key of radius $R_k=6$
  and a cylindrical cavity of radius $R_l=6$. The length of the grafted
chains is $N=10$ and their grafting density is $\sigma_g=0.01$; 
the length of free chains is $P=50$ and their bulk volume fraction is 
$\phi_b=0.5$.  Upper
panel: PMF for three different values of the Flory-Huggins interaction
parameter between free and grafted chains; lower panel: PMF for three 
different values of the Flory-Huggins interaction
parameter between free chains and particles.}
\label{fig5} 
\end{figure}
 
In all the calculations reported so far, the Flory-Huggins parameters
between all the components present in the system (free and grafted
polymer chains, solvent, and particles) have been set equal to zero,
i.e. all the effects observed in the lock-key PMFs have been purely
entropic. It would be also of interest to consider the role of
enthalpic effects in tuning lock-key interactions. To this end, we
study the effect of specific interactions between free chains and
segments comprising key and lock particles (by
varying the Flory-Huggins interaction parameter between particle
segments and free chains, $\chi_{pf}$), and  
between free and grafted chains (by
varying the Flory-Huggins interaction parameter between free and
grafted chains, $\chi_{gf}$). The former results are shown in the
lower panel of Fig.~\ref{fig5}, and the latter results in the upper
panel. In these calculations, we have set the free polymer volume
fraction at $\phi_b=0.5$, the length of free chains at $P=50$, the
grafting density at $\sigma_g=0.05$, and the sizes of key particle and
lock cavity are perfectly matched. 

All the Flory-Huggins parameters
other than those specified in the legend of Fig.~\ref{fig5} are set
equal to zero. Hence, the PMF corresponding to $\chi_{pf}$=0 in the
lower panel of Fig.~\ref{fig5} can be regarded as that of the
reference system in the absence of enthalpic effects. This reference
PMF displays weak lock-key repulsion at contact and a weak attractive
minimum at short lock-key separations.  In the presence of favorable 
 enthalpic interactions between free chains and particles  
($\chi_{pf}$=-1.5), free chains absorb on particle surfaces, 
thereby penetrating
 inside the grafted chains and making the lock-key PMF uniformly
 repulsive. When this interaction is unfavorable ($\chi_{pf}$=1.5), 
free chains are expelled from surfaces, which makes the PMF strongly
attractive. 

 Similar behavior
is observed in the upper panel of Fig.~\ref{fig5} where we vary the
enthalpic interaction between free and grafted chains. When this
interaction is favorable ($\chi_{gf}$=-0.5), the free chains wet the
brushes,\cite{egorov10b}
which produces a strongly repulsive lock-key PMF. In the opposite case   
 ($\chi_{gf}>$0), the mixing between free and grafted chains is
unfavorable, which results in a uniformly attractive lock-key PMF,
with a deep minimum at short separations. One can also note that the
range of the PMF in the upper panel is somewhat larger compared to the lower
panel, which could be due to the changes in the extension of the
grafted chains promoted by varying the $\chi_{gf}$ parameter.

\section{Conclusion}
\label{sc4}

In this work, we have carried out a mean-field theoretical study of
lock-key interactions between sterically stabilized particles in a
polymer solution. Both the key particle and the lock cavity were
assumed to have cylindrical shape, and their surfaces were uniformly
grafted with polymer chains. A two-dimensional version of 
lattice-based SCF theory was employed to compute the PMF between lock and key
particles. In analogy with earlier studies of interactions between
sterically stabilized colloids, it was found that PMF is quite
sensitive to several model parameters, making lock-key interactions
easily tunable. In particular, it was shown that when the free chains
are shorter than the grafted ones, the PMF is repulsive at all
conditions studied, while in the opposite case the lock-key
interaction can be tuned from attractive to repulsive by increasing
the volume fraction of free polymer chains. Likewise, the behavior of
the PMF, including its range, can be changed by varying the enthalpic
interactions between free chains and grafted ones or between free
chains and particles. Finally, in agreement with earlier experimental
and theoretical studies, it was observed that the strongest lock-key
interaction occurs in the case of perfect size matching between the
lock and the key; decreasing the size of the key relative to that of
the cavity weakens the interaction dramatically. 

It would be of interest to extend the present study to other
geometrical shapes of lock and key particles and to go beyond the
one-dimensional representation of the PMF by considering different
trajectories of the key approaching the lock cavity. In addition, it
would be of interest to compare the sharpness of binding-unbinding
transition (as determined by the fraction of occupied lock
cavities as a function of free chain concentration) between sterically
stabilized and charge stabilized systems.  
This will be the
subject of future research. 

\section{Acknowledgment}
The author is grateful to Prof. Frans Leermakers for
his sharing and help with using of software package $sfbox$ which was
employed to carry out some of the 2-gradient SCF calculations 
presented in this work.  


\begin{thebibliography}{10}%
\makeatletter
\providecommand \@ifxundefined [1]{%
 \ifx #1\undefined \expandafter \@firstoftwo
 \else \expandafter \@secondoftwo
\fi
}%
\providecommand \@ifnum [1]{%
 \ifnum #1\expandafter \@firstoftwo
 \else \expandafter \@secondoftwo
\fi
}%
\providecommand \enquote [1]{``#1''}%
\providecommand \bibnamefont  [1]{#1}%
\providecommand \bibfnamefont [1]{#1}%
\providecommand \citenamefont [1]{#1}%
\providecommand\href[0]{\@sanitize\@href}%
\providecommand\@href[1]{\endgroup\@@startlink{#1}\endgroup\@@href}%
\providecommand\@@href[1]{#1\@@endlink}%
\providecommand \@sanitize [0]{\begingroup\catcode`\&12\catcode`\#12\relax}%
\@ifxundefined \pdfoutput {\@firstoftwo}{%
 \@ifnum{\z@=\pdfoutput}{\@firstoftwo}{\@secondoftwo}%
}{%
 \providecommand\@@startlink[1]{\leavevmode}%
 \providecommand\@@endlink[0]{}%
}{%
 \providecommand\@@startlink[1]{%
  \leavevmode
  \pdfstartlink
   attr{/Border[0 0 1 ]/H/I/C[0 1 1]}%
   user{/Subtype/Link/A<</Type/Action/S/URI/URI(#1)>>}%
  \relax
 }%
 \providecommand\@@endlink[0]{\pdfendlink}%
}%
\providecommand \url  [0]{\begingroup\@sanitize \@url }%
\providecommand \@url [1]{\endgroup\@href {#1}{\urlprefix}}%
\providecommand \urlprefix [0]{URL }%
\providecommand \Eprint[0]{\href }%
\@ifxundefined \urlstyle {%
  \providecommand \doi [1]{doi:\discretionary{}{}{}#1}%
}{%
  \providecommand \doi [0]{doi:\discretionary{}{}{}\begingroup
  \urlstyle{rm}\Url }%
}%
\providecommand \doibase [0]{http://dx.doi.org/}%
\providecommand \Doi[1]{\href{\doibase#1}}%
\providecommand \bibAnnote [3]{%
  \BibitemShut{#1}%
  \begin{quotation}\noindent
    \textsc{Key:}\ #2\\\textsc{Annotation:}\ #3%
  \end{quotation}%
}%
\providecommand \bibAnnoteFile [2]{%
  \IfFileExists{#2}{\bibAnnote {#1} {#2} {\input{#2}}}{}%
}%
\providecommand \typeout [0]{\immediate \write \m@ne }%
\providecommand \selectlanguage [0]{\@gobble}%
\providecommand \bibinfo [0]{\@secondoftwo}%
\providecommand \bibfield [0]{\@secondoftwo}%
\providecommand \translation [1]{[#1]}%
\providecommand \BibitemOpen[0]{}%
\providecommand \bibitemStop [0]{}%
\providecommand \bibitemNoStop [0]{.\EOS\space}%
\providecommand \EOS [0]{\spacefactor3000\relax}%
\providecommand \BibitemShut [1]{\csname bibitem#1\endcsname}%
%</preamble>
\bibitem{kumar10}%
  \BibitemOpen
  \bibfield{author}{%
  \bibinfo {author} {\bibfnamefont{S.~K.}\ \bibnamefont{Kumar}}\ and\ \bibinfo
  {author} {\bibfnamefont{R.}~\bibnamefont{Krishnamoorti}},\ }%
  \bibfield{journal}{%
  \bibinfo {journal} {Annu. Rev. Chem. Biomol. Eng.}\ }%
  \textbf{\bibinfo {volume} {1}},\ \bibinfo {pages} {37} (\bibinfo {year}
  {2010})%
  \bibAnnoteFile{NoStop}{kumar10}%
\bibitem{sacanna10}%
  \BibitemOpen
  \bibfield{author}{%
  \bibinfo {author} {\bibfnamefont{S.}~\bibnamefont{Sacanna}}, \bibinfo
  {author} {\bibfnamefont{W.~T.~M.}\ \bibnamefont{Irvine}}, \bibinfo {author}
  {\bibfnamefont{P.~M.}\ \bibnamefont{Chaikin}},\ and\ \bibinfo {author}
  {\bibfnamefont{D.~J.}\ \bibnamefont{Pine}},\ }%
  \bibfield{journal}{%
  \bibinfo {journal} {Nature}\ }%
  \textbf{\bibinfo {volume} {464}},\ \bibinfo {pages} {575} (\bibinfo {year}
  {2010})%
  \bibAnnoteFile{NoStop}{sacanna10}%
\bibitem{solomon10}%
  \BibitemOpen
  \bibfield{author}{%
  \bibinfo {author} {\bibfnamefont{M.~J.}\ \bibnamefont{Solomon}},\ }%
  \bibfield{journal}{%
  \bibinfo {journal} {Nature}\ }%
  \textbf{\bibinfo {volume} {464}},\ \bibinfo {pages} {496} (\bibinfo {year}
  {2010})%
  \bibAnnoteFile{NoStop}{solomon10}%
\bibitem{kinoshita02}%
  \BibitemOpen
  \bibfield{author}{%
  \bibinfo {author} {\bibfnamefont{M.}~\bibnamefont{Kinoshita}}\ and\ \bibinfo
  {author} {\bibfnamefont{T.}~\bibnamefont{Oguni}},\ }%
  \bibfield{journal}{%
  \bibinfo {journal} {Chem.~Phys.~Lett.}\ }%
  \textbf{\bibinfo {volume} {351}},\ \bibinfo {pages} {79} (\bibinfo {year}
  {2002})%
  \bibAnnoteFile{NoStop}{kinoshita02}%
\bibitem{silvestre04}%
  \BibitemOpen
  \bibfield{author}{%
  \bibinfo {author} {\bibfnamefont{N.~M.}\ \bibnamefont{Silvestre}}, \bibinfo
  {author} {\bibfnamefont{P.}~\bibnamefont{Patricio}},\ and\ \bibinfo {author}
  {\bibfnamefont{M.~M.~T.}\ \bibnamefont{da~Gama}},\ }%
  \bibfield{journal}{%
  \bibinfo {journal} {Phys.~Rev.~E}\ }%
  \textbf{\bibinfo {volume} {69}},\ \bibinfo {pages} {061402} (\bibinfo {year}
  {2004})%
  \bibAnnoteFile{NoStop}{silvestre04}%
\bibitem{konig08}%
  \BibitemOpen
  \bibfield{author}{%
  \bibinfo {author} {\bibfnamefont{P.~M.}\ \bibnamefont{Konig}}, \bibinfo
  {author} {\bibfnamefont{R.}~\bibnamefont{Roth}},\ and\ \bibinfo {author}
  {\bibfnamefont{S.}~\bibnamefont{Dietrich}},\ }%
  \bibfield{journal}{%
  \bibinfo {journal} {Europhys. Lett.}\ }%
  \textbf{\bibinfo {volume} {84}},\ \bibinfo {pages} {5} (\bibinfo {year}
  {2008})%
  \bibAnnoteFile{NoStop}{konig08}%
\bibitem{odriozola08}%
  \BibitemOpen
  \bibfield{author}{%
  \bibinfo {author} {\bibfnamefont{G.}~\bibnamefont{Odriozola}}, \bibinfo
  {author} {\bibfnamefont{F.}~\bibnamefont{Jimenez-Angeles}},\ and\ \bibinfo
  {author} {\bibfnamefont{M.}~\bibnamefont{Lozada-Cassou}},\ }%
  \bibfield{journal}{%
  \bibinfo {journal} {J.~Chem.~Phys.}\ }%
  \textbf{\bibinfo {volume} {129}},\ \bibinfo {pages} {111101} (\bibinfo {year}
  {2008})%
  \bibAnnoteFile{NoStop}{odriozola08}%
\bibitem{jin11}%
  \BibitemOpen
  \bibfield{author}{%
  \bibinfo {author} {\bibfnamefont{Z.}~\bibnamefont{Jin}}\ and\ \bibinfo
  {author} {\bibfnamefont{J.~Z.}\ \bibnamefont{Wu}},\ }%
  \bibfield{journal}{%
  \bibinfo {journal} {J.~Phys.~Chem.~B}\ }%
  \textbf{\bibinfo {volume} {115}},\ \bibinfo {pages} {1450} (\bibinfo {year}
  {2011})%
  \bibAnnoteFile{NoStop}{jin11}%
\bibitem{asakura54}%
  \BibitemOpen
  \bibfield{author}{%
  \bibinfo {author} {\bibfnamefont{S.}~\bibnamefont{Asakura}}\ and\ \bibinfo
  {author} {\bibfnamefont{F.}~\bibnamefont{Oosawa}},\ }%
  \bibfield{journal}{%
  \bibinfo {journal} {J.~Chem.~Phys.}\ }%
  \textbf{\bibinfo {volume} {22}},\ \bibinfo {pages} {1255} (\bibinfo {year}
  {1954})%
  \bibAnnoteFile{NoStop}{asakura54}%
\bibitem{fleer93}%
  \BibitemOpen
  \bibfield{author}{%
  \bibinfo {author} {\bibfnamefont{G.~J.}\ \bibnamefont{Fleer}}, \bibinfo
  {author} {\bibfnamefont{M.~A.~C.}\ \bibnamefont{Stuart}}, \bibinfo {author}
  {\bibfnamefont{J.~M. H.~M.}\ \bibnamefont{Scheutjens}}, \bibinfo {author}
  {\bibfnamefont{T.}~\bibnamefont{Cosgrove}},\ and\ \bibinfo {author}
  {\bibfnamefont{B.}~\bibnamefont{Vincent}},\ }%
  \emph{\bibinfo {title} {Polymers at Interfaces}}\ (\bibinfo {publisher}
  {Chapman and Hall},\ \bibinfo {address} {London},\ \bibinfo {year} {1993})%
  \bibAnnoteFile{NoStop}{fleer93}%
\bibitem{egorov07}%
  \BibitemOpen
  \bibfield{author}{%
  \bibinfo {author} {\bibfnamefont{A.}~\bibnamefont{Striolo}}\ and\ \bibinfo
  {author} {\bibfnamefont{S.~A.}\ \bibnamefont{Egorov}},\ }%
  \bibfield{journal}{%
  \bibinfo {journal} {J.~Chem.~Phys.}\ }%
  \textbf{\bibinfo {volume} {126}},\ \bibinfo {pages} {014902} (\bibinfo {year}
  {2007})%
  \bibAnnoteFile{NoStop}{egorov07}%
\bibitem{egorov10b}%
  \BibitemOpen
  \bibfield{author}{%
  \bibinfo {author} {\bibfnamefont{A.}~\bibnamefont{Milchev}}, \bibinfo
  {author} {\bibfnamefont{S.~A.}\ \bibnamefont{Egorov}},\ and\ \bibinfo
  {author} {\bibfnamefont{K.}~\bibnamefont{Binder}},\ }%
  \bibfield{journal}{%
  \bibinfo {journal} {J.~Chem.~Phys.}\ }%
  \textbf{\bibinfo {volume} {132}},\ \bibinfo {pages} {184905} (\bibinfo {year}
  {2010})%
  \bibAnnoteFile{NoStop}{egorov10b}%
\bibitem{dutta08}%
  \BibitemOpen
  \bibfield{author}{%
  \bibinfo {author} {\bibfnamefont{N.}~\bibnamefont{Dutta}}\ and\ \bibinfo
  {author} {\bibfnamefont{D.}~\bibnamefont{Green}},\ }%
  \bibfield{journal}{%
  \bibinfo {journal} {Langmuir}\ }%
  \textbf{\bibinfo {volume} {24}},\ \bibinfo {pages} {5260} (\bibinfo {year}
  {2008})%
  \bibAnnoteFile{NoStop}{dutta08}%
\bibitem{green06}%
  \BibitemOpen
  \bibfield{author}{%
  \bibinfo {author} {\bibfnamefont{D.~L.}\ \bibnamefont{Green}}\ and\ \bibinfo
  {author} {\bibfnamefont{J.}~\bibnamefont{Mewis}},\ }%
  \bibfield{journal}{%
  \bibinfo {journal} {Langmuir}\ }%
  \textbf{\bibinfo {volume} {22}},\ \bibinfo {pages} {9546} (\bibinfo {year}
  {2006})%
  \bibAnnoteFile{NoStop}{green06}%
\bibitem{bansal05}%
  \BibitemOpen
  \bibfield{author}{%
  \bibinfo {author} {\bibfnamefont{A.}~\bibnamefont{Bansal}}, \bibinfo {author}
  {\bibfnamefont{H.~C.}\ \bibnamefont{Yang}}, \bibinfo {author}
  {\bibfnamefont{C.~Z.}\ \bibnamefont{Li}}, \bibinfo {author}
  {\bibfnamefont{B.~C.}\ \bibnamefont{Benicewicz}}, \bibinfo {author}
  {\bibfnamefont{S.~K.}\ \bibnamefont{Kumar}},\ and\ \bibinfo {author}
  {\bibfnamefont{L.~S.}\ \bibnamefont{Schadler}},\ }%
  \bibfield{journal}{%
  \bibinfo {journal} {Nat. Mater.}\ }%
  \textbf{\bibinfo {volume} {4}},\ \bibinfo {pages} {693} (\bibinfo {year}
  {2005})%
  \bibAnnoteFile{NoStop}{bansal05}%
\bibitem{bansal06}%
  \BibitemOpen
  \bibfield{author}{%
  \bibinfo {author} {\bibfnamefont{A.}~\bibnamefont{Bansal}}, \bibinfo {author}
  {\bibfnamefont{H.~C.}\ \bibnamefont{Yang}}, \bibinfo {author}
  {\bibfnamefont{C.~Z.}\ \bibnamefont{Li}}, \bibinfo {author}
  {\bibfnamefont{B.~C.}\ \bibnamefont{Benicewicz}}, \bibinfo {author}
  {\bibfnamefont{S.~K.}\ \bibnamefont{Kumar}},\ and\ \bibinfo {author}
  {\bibfnamefont{L.~S.}\ \bibnamefont{Schadler}},\ }%
  \bibfield{journal}{%
  \bibinfo {journal} {J. Poly. Sci. B: Polym. Phys.}\ }%
  \textbf{\bibinfo {volume} {44}},\ \bibinfo {pages} {2944} (\bibinfo {year}
  {2006})%
  \bibAnnoteFile{NoStop}{bansal06}%
\bibitem{huinink97}%
  \BibitemOpen
  \bibfield{author}{%
  \bibinfo {author} {\bibfnamefont{H.~P.}\ \bibnamefont{Huinink}}, \bibinfo
  {author} {\bibfnamefont{A.}~\bibnamefont{de~Keizer}}, \bibinfo {author}
  {\bibfnamefont{F.~A.~M.}\ \bibnamefont{Leermakers}},\ and\ \bibinfo {author}
  {\bibfnamefont{J.}~\bibnamefont{Lyklema}},\ }%
  \bibfield{journal}{%
  \bibinfo {journal} {Langmuir}\ }%
  \textbf{\bibinfo {volume} {13}},\ \bibinfo {pages} {6618} (\bibinfo {year}
  {1997})%
  \bibAnnoteFile{NoStop}{huinink97}%
\bibitem{steels00}%
  \BibitemOpen
  \bibfield{author}{%
  \bibinfo {author} {\bibfnamefont{B.~M.}\ \bibnamefont{Steels}}, \bibinfo
  {author} {\bibfnamefont{F.~A.~M.}\ \bibnamefont{Leermakers}},\ and\ \bibinfo
  {author} {\bibfnamefont{C.~A.}\ \bibnamefont{Haynes}},\ }%
  \bibfield{journal}{%
  \bibinfo {journal} {J. Chromatography B}\ }%
  \textbf{\bibinfo {volume} {743}},\ \bibinfo {pages} {31} (\bibinfo {year}
  {2000})%
  \bibAnnoteFile{NoStop}{steels00}%
\bibitem{steels00b}%
  \BibitemOpen
  \bibfield{author}{%
  \bibinfo {author} {\bibfnamefont{B.~M.}\ \bibnamefont{Steels}}, \bibinfo
  {author} {\bibfnamefont{J.}~\bibnamefont{Koska}},\ and\ \bibinfo {author}
  {\bibfnamefont{C.~A.}\ \bibnamefont{Haynes}},\ }%
  \bibfield{journal}{%
  \bibinfo {journal} {J. Chromatography B}\ }%
  \textbf{\bibinfo {volume} {743}},\ \bibinfo {pages} {41} (\bibinfo {year}
  {2000})%
  \bibAnnoteFile{NoStop}{steels00b}%
\bibitem{vanlent90}%
  \BibitemOpen
  \bibfield{author}{%
  \bibinfo {author} {\bibfnamefont{B.}~\bibnamefont{van Lent}}, \bibinfo
  {author} {\bibfnamefont{R.}~\bibnamefont{Israels}}, \bibinfo {author}
  {\bibfnamefont{J.~M. H.~M.}\ \bibnamefont{Scheutjens}},\ and\ \bibinfo
  {author} {\bibfnamefont{G.~J.}\ \bibnamefont{Fleer}},\ }%
  \bibfield{journal}{%
  \bibinfo {journal} {J. Coll. Interface Sci.}\ }%
  \textbf{\bibinfo {volume} {137}},\ \bibinfo {pages} {380} (\bibinfo {year}
  {1990})%
  \bibAnnoteFile{NoStop}{vanlent90}%
\bibitem{trombly10}%
  \BibitemOpen
  \bibfield{author}{%
  \bibinfo {author} {\bibfnamefont{D.~M.}\ \bibnamefont{Trombly}}\ and\
  \bibinfo {author} {\bibfnamefont{V.}~\bibnamefont{Ganesan}},\ }%
  \bibfield{journal}{%
  \bibinfo {journal} {J.~Chem.~Phys.}\ }%
  \textbf{\bibinfo {volume} {133}},\ \bibinfo {pages} {154904} (\bibinfo {year}
  {2010})%
  \bibAnnoteFile{NoStop}{trombly10}%
\end{thebibliography}
\end{document}